\documentclass[
    ,final            
  ]
  {aipproc}

\usepackage{epsfig}
\layoutstyle{6x9}

\begin{document}
\begin{flushright}
BNL-NT-02/25\\
RBRC-295\\
\vspace*{-0.9cm}
\end{flushright}

\title{NLO QCD Corrections to $\mathbf{A_{\mathrm{LL}}^{\pi}}$}

\author{Barbara J\"{a}ger}{address={Inst.\ for Theor.\ Physics, Univ.\ of Regensburg, 
D-93040 Regensburg, Germany}}

\author{\underline{Marco Stratmann}}{address={Inst.\ for Theor.\ Physics, Univ.\ of Regensburg, 
D-93040 Regensburg, Germany}}

\author{Werner Vogelsang}{address={RBRC and Physics Department,
Brookhaven National Laboratory, Upton, NY 11973, U.S.A.}}

\begin{abstract}
We present a calculation for single-inclusive large-$p_T$ pion production in 
longitudinally polarized $pp$ collisions in next-to-leading order QCD. 
The corresponding double-spin asymmetry $A_{\mathrm{LL}}^{\pi}$ 
for this process will soon be used at BNL-RHIC to measure $\Delta g$.
\end{abstract}

\maketitle

\section{Theoretical Framework}
%
Very inelastic $pp$ collisions with longitudinally polarized beams at the BNL-RHIC will 
open up unequaled possibilities to measure the so far elusive 
polarized gluon density $\Delta g$.  
RHIC has the advantage of operating at high energies ($\sqrt{S}=200$ and $500$ GeV),
where the underlying theoretical framework, i.e., perturbative QCD, is 
expected to be under good control. 
In  addition, it offers various different channels in which $\Delta g$ can be studied, 
such as prompt-$\gamma$, heavy flavor, jet or inclusive-hadron production 
\cite{ref:rhicreport,ref:spintalk}. In this way, RHIC will provide the best 
source of information on $\Delta g$ for a long time to come.

The basic concept that underlies most of spin physics at RHIC
is the factorization theorem. It states that large
momentum-transfer reactions may be factorized at a scale $\mu_F$ into
long-distance pieces that contain the desired information on the
spin structure of the nucleon in terms of its {\em universal} parton
densities, such as $\Delta g$, and parts that are
short-distance and describe the hard interactions of the
partons. The latter can be evaluated using perturbative QCD.
The factorization scale $\mu_F$ is not further specified by the theory but usually
chosen to be of the order of the hard scale in the reaction.

In the following, we consider the spin-dependent cross section
\begin{equation} 
\label{eq:eq1}
d\Delta \sigma \equiv
\frac{1}{2} \left[d\sigma^{++} - d\sigma^{+-}\right] \;\; ,
\end{equation}
where the superscripts denote the helicities of the 
protons in the scattering, for the reaction $pp\to \pi X$, where the pion is at high
transverse momentum $p_T$, ensuring large momentum transfer. 
The statement of the factorization theorem is then
\begin{eqnarray}
\label{eq:eq2}
d\Delta \sigma &=&\sum_{a,b,c}\, 
\int dx_a 
\int dx_b 
\int dz_c \,\,
\Delta f_a (x_a,\mu_F) \,\Delta f_b (x_b,\mu_F) 
D_c^{\pi}(z_c,\mu_F') \, \nonumber \\ [2mm]
&&
\times \, d\Delta \hat{\sigma}_{ab}^{c}
(x_aP_A, x_bP_B, P_{\pi}/z_c, \mu_R, \mu_F, \mu_F') \;\; ,
\end{eqnarray}
where the sum is over all  contributing partonic channels $a+b\to c + X$, with
$d\Delta \hat{\sigma}_{ab}^{c}$ the associated partonic cross
section, defined in complete analogy with Eq.~(\ref{eq:eq1}).
Besides the factorization scale $\mu_F$ for the initial-state partons
$\Delta f_{a,b}$, there is also a factorization scale $\mu_F'$ for the absorption of 
long-distance effects into the parton-to-pion fragmentation functions $D_c^{\pi}$. 
The renormalization scale $\mu_R$ in (\ref{eq:eq2}) is associated with the running of $\alpha_s$.

It is planned for the coming RHIC run (early 2003) to attempt a first measurement of 
\begin{equation}
\label{eq:eq3}
A_{LL}^{\pi}=\frac{d\Delta \sigma}{d\sigma}=\frac{ 
d\sigma^{++} - d\sigma^{+-}}{d\sigma^{++} + d\sigma^{+-}}
\end{equation}
for high-$p_T$ pion production. The main underlying idea here is that
the spin asymmetry $A_{LL}^{\pi}$ is very sensitive to $\Delta g$ through 
the contributions from polarized quark-gluon and gluon-gluon scatterings. 
In general, a leading-order (LO) estimate of (\ref{eq:eq2}) or (\ref{eq:eq3}) 
merely captures the main features, but does not usually provide a quantitative understanding. 
For instance, the dependence on the unphysical scales $\mu_F$, $\mu_F'$, and
$\mu_R$ is expected to be much reduced when going to higher orders in the perturbative
expansion. Hence, only with knowledge of the next-to-leading order (NLO)
QCD corrections can one reliably  extract information on the parton distribution functions 
from the reaction. A NLO calculation of $A_{LL}^{\pi}$ has been completed very recently
\cite{ref:nlohadron}, and here we briefly sketch the results; for details, see 
\cite{ref:nlohadron}. We note that the PHENIX collaboration has recently 
presented first, still preliminary, results for the 
{\em un}polarized cross section for $pp\to \pi^0 X$ at $\sqrt{S}=200$~GeV, 
which are well described by a NLO QCD calculation~\cite{ref:phenix}.

The partonic cross sections $d\Delta \hat{\sigma}_{ab}^{c}$ in (\ref{eq:eq2})
have to be summed over all final states (excluding $c$ which fragments)
and integrated over the entire phase space of $X$. 
The LO results, which have been known for a long time \cite{ref:lo},
are obtained from evaluating all tree-level $2\to 2$ QCD scattering diagrams.
At NLO, we have ${\cal O}(\alpha_s)$ corrections to the LO 
reactions, and also additional new processes, giving rise to 16
different channels in total, like $qq \rightarrow q X$, $qg \rightarrow g X$, etc.
At intermediate stages the NLO calculation will necessarily show singularities
that represent the long-distance sensitivity. In addition,
for those processes that are already present at LO,
real $2\to 3$ and virtual one-loop $2\to 2$ contributions 
will individually have infrared (IR) singularities that only cancel in their sum. 
Virtual diagrams will also produce ultraviolet (UV)
poles that need to be removed by the renormalization of the
strong coupling constant at a scale $\mu_R$.
We choose $n=4-2\varepsilon$ dimensional regularization to make these singularities manifest. 
Subtractions of poles will generally be made in the $\overline{\rm{MS}}$ scheme.
We use the HVBM prescription \cite{ref:hvbm} to describe polarizations of 
particles in $n$ dimensions.

At ${\cal O}(\alpha_s^3)$, virtual corrections, which we have calculated adopting
two different methods, only contribute through their interference with the Born diagrams.
Firstly, one could make use of known $\overline{\rm{MS}}$-renormalized 
one-loop vertex and self-energy insertions as given in \cite{ref:nps}. 
Only the UV-finite box diagrams have to be calculated from scratch.
The second approach makes use of the fact that helicity amplitudes for 
all one-loop $2\to 2$ QCD scattering diagrams were presented in \cite{ref:kst}.  
These results will not immediately yield the answer for the HVBM prescription 
but the transformation is straightforward.

In the $2\to 3$ contributions, the two unobserved partons 
need to be integrated over their entire phase space which we perform
{\em analytically}. In this way the final answer is much more amenable
to a numerical evaluation, giving stable results in a short time. 
This may become important when experimental data will 
become available, and one is aiming to extract $\Delta g$ from 
them within a ``global analysis'' \cite{ref:sv}.
Phase space integrations are organized best in the rest frame of the 
two unobserved partons. Extensive partial fractioning of the matrix elements 
then always leads to a ``master integral'' which can be done analytically.
Singularities when the invariant mass of the unobserved partons vanishes
are made manifest with help of the usual ``$+$''-distributions.

All genuine IR singularities cancel in the sum of all contributions.
However, the limit $\varepsilon \to 0$ still cannot be taken 
as a result of collinear divergencies. These remaining poles need to be 
factored into the bare parton distribution and fragmentation functions, 
depending on whether their origin was in the initial or final state. 
This standard procedure introduces the factorization scales 
$\mu_F$ and $\mu_F'$ in Eq.~(\ref{eq:eq2}).
We note that we have simultaneously computed also the NLO corrections for
the unpolarized case, where we fully agree at an {\em analytical level}
with results available in the literature \cite{ref:unpolnlo}. This 
provides an extremely powerful check on the correctness of all our
calculations.

Finally, we note that the same NLO calculation was presented in \cite{ref:daniel}
based on MC phase space integration techniques. Such an approach has the advantage
of being very flexible as it may be used for any IR-safe observable, with any
experimental cut. However, the numerical integrations are delicate and time-consuming. 
Early comparisons show very good agreement of the numerical results.

\section{Numerical Results}
%
For our numerical calculations we assume the same kinematic coverage as 
in the unpolarized PHENIX measurement mentioned above \cite{ref:phenix}:
$\sqrt{S}=200\,\mathrm{GeV}$, pion transverse momenta 
in the range $2\leq p_T \leq 13$ GeV, and pseudorapidities integrated over $|\eta|\leq 0.38$. 
We also always take into account that the pion measurement is at present possible 
only over half the azimuthal angle. 
To calculate the NLO/LO polarized cross section (\ref{eq:eq2}) we use
the spin-dependent GRSV parton densities (``GRSV-std'') and the
pion fragmentation functions of \cite{ref:kkp}.
To investigate the sensitivity of $A_{LL}^{\pi}$ to $\Delta g$, we also use a set, 
for which $\Delta g$ is assumed to be particularly large (``GRSV-max''). 
For the NLO (LO) unpolarized cross section, 
we use the CTEQ5M (CTEQ5L)~\cite{ref:cteq5m} densities.
\begin{figure}[t]
\epsfig{figure=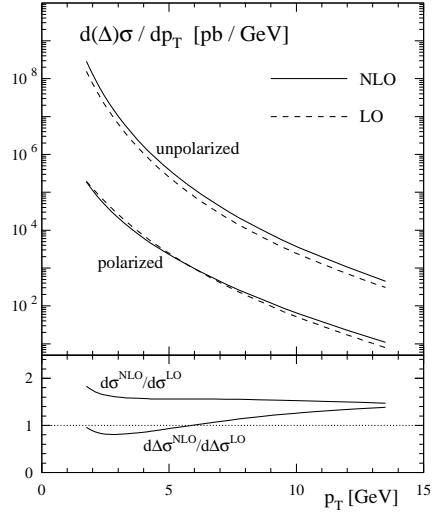,width=0.45\textwidth}
\caption{Unpolarized and polarized $\pi^0$ production cross sections in 
NLO (solid) and LO (dashed) at $\sqrt{S}=200$ GeV. The lower panel 
shows the $K$-factor in each case. Figure taken from \cite{ref:nlohadron}.  \label{fig4}}
\end{figure}
%
\begin{figure}[hb]
\epsfig{figure=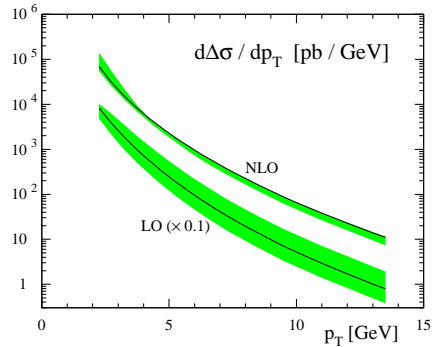,width=0.45\textwidth}
\caption{Scale dependence of the polarized cross section
for $\pi^0$ production at LO and NLO \cite{ref:nlohadron} in the range 
$p_T/2 \leq \mu_R=\mu_F=\mu_F' \leq 2 p_T$. We have rescaled the LO results by $0.1$ to 
separate them better from the NLO ones. In each case the solid line corresponds
to the choice where all scales are set to $p_T$.\label{fig5}}
\end{figure}

Figure~\ref{fig4} shows our results for the unpolarized and polarized
cross sections at NLO and LO, where we have chosen the scales
$\mu_R=\mu_F=\mu_F'=p_T$. The lower part of the figure displays
the ``$K$-factor'', $K=d(\Delta)\sigma^{\mathrm{NLO}}/d(\Delta)\sigma^{\mathrm{LO}}$.
One can see that in the unpolarized case the corrections are roughly 
constant and about $50\%$ over the $p_T$-region considered. 
In the polarized case, we find
generally smaller corrections which become of similar size as those
for the unpolarized case only at the high-$p_T$ end. The cross section for
$p_T$-values smaller than about 2 GeV is outside the domain of 
perturbative calculations as indicated by rapidly increasing
NLO corrections and, therefore, is not considered here.

Figure~\ref{fig5} shows the improvement in scale dependence of the spin-dependent
cross section when going from LO to NLO. 
In each case the shaded bands indicate the uncertainties from varying the
unphysical scales in the range $p_T/2 \leq \mu_R=\mu_F=\mu_F' \leq 2 p_T$. The solid
lines are for the choice where all scales are set to $p_T$.
One can see that the scale dependence indeed becomes much smaller at NLO.
 
\begin{figure}[t]
\epsfig{figure=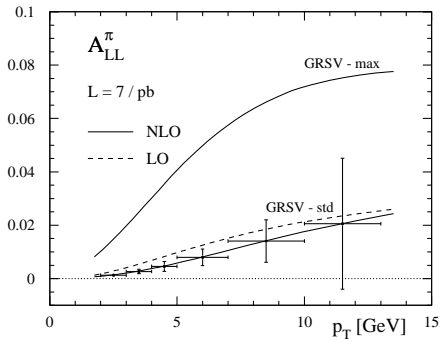,width=0.45\textwidth}
\caption{Spin asymmetry for $\pi^0$ production in NLO (solid lines).
The dashed line shows the asymmetry at LO for the GRSV ``standard''
set. The ``error bars'' indicate the expected statistical accuracy targeted
for the upcoming run of RHIC (see text). Figure taken from \cite{ref:nlohadron}. 
\label{fig6}}
\end{figure}

Results for $A_{LL}^{\pi}$ are given in Fig.~\ref{fig6}.
We have again chosen all scales to be $p_T$. As expected from the larger 
$K$-factor for the unpolarized cross section shown in Fig.~\ref{fig4}, the 
asymmetry is somewhat smaller at NLO than at LO, showing that
inclusion of NLO QCD corrections is rather important for the
analysis of the data in terms of $\Delta g$. 

%
%
We also conclude from the figure that there are excellent prospects for 
determining $\Delta g(x)$ from $A_{LL}^{\pi}$ measurements at
RHIC: the asymmetries found for the two different sets of polarized 
parton densities, which mainly differ in the gluon density, show
marked differences, much larger than the expected statistical
errors in the experiment, indicated in the figure. The latter
may be estimated by the formula
$\delta A_{LL}^{\pi} = 1/(P^2\sqrt{{\cal L}\sigma_{\rm bin}})$,
where $P$ is the polarization of one beam, ${\cal L}$ the integrated
luminosity, and $\sigma_{\rm bin}$ the unpolarized
cross section integrated over the $p_T$-bin for which the error is to be
determined. We have used very moderate values $P=0.4$ and
${\cal L}=7$/pb, which are targets for the coming run. 

To conclude, we have presented the results of a largely analytical 
computation of the NLO partonic hard-scattering cross sections relevant for the
spin asymmetry $A_{LL}^{\pi}$ for high-$p_T$ pion production 
in longitudinally polarized hadron-hadron collisions. 
The asymmetry turns out to be a promising
tool to provide first information on $\Delta g$
even for the rather moderate luminosities targeted for the
coming run with polarized protons at RHIC.

B.J.\ is supported by the European Commission IHP program under contract
HPRN-CT-2000-00130. W.V.\ is grateful to RIKEN, Brookhaven National Laboratory and the U.S.\
Department of Energy (contract number DE-AC02-98CH10886) for
providing the facilities essential for the completion of this work.


%

\begin{thebibliography}{99}
%
\bibitem{ref:rhicreport} For a review on RHIC spin, see: G.\ Bunce et al.,
{\it Annu. Rev. Nucl. Part. Sci.} {\bf 50}, 525 (2000).
%
\bibitem{ref:spintalk} M.\ Stratmann, these proceedings.
%
\bibitem{ref:nlohadron} B.\ J\"{a}ger, A.\ Sch\"{a}fer, M.\ Stratmann, and W.\ Vogelsang, 
{\tt hep-ph/0211007}.
%
\bibitem{ref:phenix} H.\ Torii, talk presented at 
{\em Quark Matter 2002}, Nantes, France, 2002.
%
\bibitem{ref:lo} See, for example: J.\ Babcock et al.,
{\it Phys. Rev. Lett.} {\bf 40}, 1161 (1978); {\it Phys. Rev.} {\bf D19}, 1483 (1979);
R.\ Gastmans and T.T.\ Wu, {\em The ubiquitous photon}, Clarendon Press, Oxford, 1990. 
%
\bibitem{ref:hvbm} G.\ 't Hooft and M.\ Veltman, {\it Nucl. Phys.} {\bf B44}, 189 (1972);
P.\ Breitenlohner and D.\ Maison, {\it Commun. Math. Phys.} {\bf 52}, 11 (1977).
%
\bibitem{ref:nps} M.A.\ Nowak, M.\ Praszalowicz, and W.\ Slominski,
{\it Annals Phys.} {\bf 166}, 443 (1986).
%
\bibitem{ref:kst} Z.\ Kunszt, A.\ Signer, and Z.\ Trocsanyi,
{\it Nucl. Phys.} {\bf B411}, 397 (1994).
%
\bibitem{ref:sv} M.\ Stratmann and W.\ Vogelsang, {\it Phys. Rev.} {\bf D64}, 114007 (2001). 
%
\bibitem{ref:unpolnlo} R.K.\ Ellis and J.C.\ Sexton, {\it Nucl. Phys.} {\bf B269}, 445 (1986);
F.\ Aversa et al.,
{\it ibid.} {\bf B327}, 105 (1989).
%
\bibitem{ref:daniel} D. de Florian, {\tt hep-ph/0210442}.
%
\bibitem{ref:grsv}  M.\ Gl\"{u}ck, E.\ Reya, M.\ Stratmann, and
W.\ Vogelsang, {\em Phys. Rev.} {\bf D63}, 094005 (2001). 
%
\bibitem{ref:kkp}  B.A.\ Kniehl, G.\ Kramer, and B.\ P\"{o}tter,
{\em Nucl. Phys.} {\bf B582}, 514 (2000). 
%
\bibitem{ref:cteq5m} CTEQ Collaboration, H.-L. Lai et al., 
{\em Eur. Phys. J.} {\bf C12}, 375 (2000).
%
\end{thebibliography}
\end{document}